\documentclass[12pt,preprint]{aastex}

\def\be{\begin{equation}}
\def\ee{\end{equation}}

\def\msun{M_{\odot}}

\def\be{\begin{equation}}
\def\ee{\end{equation}}
\catcode`\@=11 
\def\@versim#1#2{\vcenter{\offinterlineskip
        \ialign{$\m@th#1\hfil##\hfil$\crcr#2\crcr\sim\crcr } }}
\def\lsim{\mathrel{\mathpalette\@versim<}}
\def\gsim{\mathrel{\mathpalette\@versim>}}

\shorttitle{LHAF: thermal equilibrium curve and thermal stability}
\shortauthors{Feng Yuan}

\begin{document}

\title{Luminous hot accretion flows: thermal equilibrium curve
and thermal stability}

\author{Feng Yuan}
\affil{Harvard-Smithsonian Center for Astrophysics, 60 Garden Street,
Cambridge, MA 02138; fyuan@cfa.harvard.edu}


\begin{abstract}
In a previous paper, we presented the global solution of a new 
accretion flow model, namely luminous hot accretion flows (LHAFs). 
In this {\em Letter},
we first show the corresponding thermal equilibrium curve 
of LHAFs in the mass accretion rate vs.
surface density diagram. Then we examine its thermal stability
again local perturbations. We find that LHAFs are thermally unstable
when thermal conduction is neglected. 
However, when the accretion rate is not very large, the timescale of 
the growth of perturbations is longer than the accretion 
timescale, therefore the instability has no dynamical effect on the 
accretion flow. When the accretion rate is large, the perturbations
can grow very fast at a certain radius. As a result,
some cold clumps may form and the accretion 
flow will become multi-phase.
\end{abstract}



\section{Introduction: the physics of the new hot accretion disk solution}

There has been a great interest to the accretion process around black holes. 
The most famous accretion solution is the geometrically thin and
optically thick cold disk model 
developed by Shakura \& Sunyaev (1973; hereafter SSD) and others.
The second solution was discovered by Shapiro, Lightman \& Eardley (1976;
hereafter SLE). This solution is optically thin
and hot, with $T_e \sim 10^9$K. Pringle (1976) found that SLE
is thermally unstable although it is not clear what is the consequence of such
instability. 

In both SSD and SLE, the energy advection is neglected.
It was found that when the mass accretion rate is higher than 
the Eddington rate $\dot{M}_{\rm Edd} (\equiv 10L_{\rm Edd}/c^2)$, the large
optical depth traps most of the photons therefore most of the viscously dissipated
energy is stored in the gas and advected into the black hole rather
than radiated away (``slim disk''; Abramowicz et al. 1988).

The fourth accretion solution is the advection-dominated accretion flow (ADAF;
Narayan \& Yi 1995; Abramowicz et al.\ 1995, hereafter A95;
see reviews by Narayan, Mahadevan \& Quataert 1998; Kato, Fukue \&
Mineshige 1998). 
Different from SLE,
the energy advection is included in the ions energy equation of an ADAF,
$Q_{\rm adv}=Q_{\rm vis}-Q_{\rm ie}$. Here $Q_{\rm adv}, Q_{\rm vis}$ and
$Q_{\rm ie}$ are the rates of energy advection, viscous heating and Coulomb
cooling per unit area of the accretion disk, respectively.
In a typical ADAF, the density of gas is very low, so $Q_{\rm ie}\ll
Q_{\rm vis}\approx Q_{\rm adv}$, i.e., {\em the viscous heating is balanced by
advective cooling}. Since $Q_{\rm ie} \propto \dot{M}^2$ 
while $Q_{\rm vis} \propto \dot{M}$, i.e., $Q_{\rm ie}$ increases
faster than $Q_{\rm vis}$ with increasing $\dot{M}$, there exists a critical
rate $\dot{M}_1$, determined by $Q_{\rm vis}\approx Q_{\rm ie}$.
At this rate, a large fraction of the viscously dissipated energy is 
transferred to the electrons and radiated away, so the
accretion flow ceases to be advection-dominated. 
Above $\dot{M}_1$, it was thought previously that no {\em hot}
accretion solution exists and the only viable solution is SSD.

However, our recent work (Yuan 2001, hereafter Y01) indicated
that this is not true:
above $\dot{M}_1$, a new hot accretion solution exists.
To illustrate, let's first write out the
formula of the energy advection of ions,
\be
Q_{\rm adv}=
\rho H v_r\left[\frac{k}{\mu m_{\mu}}\frac{1}{\gamma-1}
\frac{dT}{dr}-\frac{kT}{\mu m_{\mu}}\frac{1}{\rho}\frac{d\rho}{dr}\right]
\equiv Q_{\rm int} - Q_{\rm com},
\ee
i.e., the energy advection consists of two terms, namely the 
internal energy gradient term $Q_{\rm int}$ and the
compression work $Q_{\rm com}$. 
{\em The full condition for the existence of a hot accretion solution, when the 
flow starts out hot, is $dT/dr<0$, i.e., $Q_{\rm int}>0$} ({\em note $v_r <0$}).  
For an ADAF, $\dot{M}<\dot{M}_1$, $Q_{\rm adv}=Q_{\rm vis}-Q_{ie}>0$,
so $Q_{\rm int}=Q_{\rm adv}+Q_{\rm com}>0$. 
When $\dot{M}>\dot{M}_1$, $Q_{\rm adv}=Q_{\rm vis}-Q_{ie}<0$, but there
obviously exists another critical accretion rate, $\dot{M}_2$, determined by
$Q_{\rm com}+Q_{\rm vis} \approx Q_{ie}$. Below $\dot{M}_2$, we still have
$Q_{\rm int}=Q_{\rm vis}+Q_{\rm com}-Q_{ie}>0$. 
Therefore, the accretion flow can remain hot if it 
starts out hot. We denote this new solution as
luminous hot accretion flow (LHAF hereafter). 

From the above analysis, we see that 
LHAFs are along the line of ADAFs---the equations describing both are completely
the same and we just extend ADAFs to higher accretion rates. However, the dynamics
of an LHAF is quite different from an ADAF.
In an ADAF, $Q_{\rm adv}>0$,
i.e., the advection plays a ``cooling'' role from the Lagrangian point
of view. But in an LHAF, $Q_{\rm adv}<0$, so it plays a ``heating''
role. In the language of entropy, it is the conversion
of entropy together with viscous dissipation that 
supplies the radiation of LHAFs. In this sense, 
an LHAF is dynamically similar with Bondi accretion and cooling flow in
galactic clusters. 

In this context we can understand why previous authors didn't
find this solution. In Narayan \& Yi (1995) and 
Esin et al. (1997), 
they {\it a priori} set the advection factor
``$f$''($\equiv Q_{\rm adv}/Q_{\rm vis}$)
as positive, so they only obtained the ADAF solution.
Similarly, A95 and Chen et al. (1995, hereafter C95) didn't find this solution
because they set the parameter $\xi$ (see eq. (6) below) as positive.


In this {\it Letter}, we will first show the thermal equilibria of LHAFs  
in the $\dot{M}$ vs. $\Sigma$ (surface density) diagram (\S 2). Such a diagram 
is widely used in the study of 
accretion disks (e.g., A95; C95;
Kusunose \& Mineshige 1996; Bj\"onsson et al. 1996). Then
in \S3 we investigate the thermal stability of LHAFs.

\section{Thermal equilibrium curve of LHAFs}

We take a one-temperature accretion flow as an example for simplicity.
For the purpose of comparison, we adopt almost exactly the same
equations as C95 (see also A95). 

\be
\dot{M}=-2\pi R \Sigma v_r,
\ee
\be
\nu \Sigma = \frac{\dot{M}}{3\pi}f_* g^{-1},
\ee
\be
Q_{\rm adv}=Q_{\rm vis}-Q_{\rm rad}.
\ee
They represent conservations of mass, angular 
momentum, and energy, respectively.
Here $\nu=\frac{2}{3}\alpha c_s H$ is the kinetic viscosity coefficient,
$\Sigma=2\rho H$ is the surface density, $f_*=1-9\Omega(3R_s)/
[\Omega(R)(R/R_s)^2]$ with $R_s=2GM/c^2, g=-\frac{2}{3}
(d{\rm ln}\Omega/d{\rm ln}R)$.
The forms of $Q_{\rm vis}, Q_{\rm adv}$ and $Q_{\rm rad}$ are,
\be
Q_{\rm vis}=\frac{3\dot{M}}{4\pi}\Omega^2f_*g,
\ee
\be
Q_{\rm adv}=-\Sigma v_r \frac{p}{\rho}\xi = \frac{\dot{M}}
{2 \pi R^2}\frac{p}{\rho}\xi,
\ee
\be
Q_{\rm rad}=8\sigma T^4\left(\frac{3\tau}{2}+\sqrt{3}+\frac{8\sigma T^4}
{Q_{\rm brem}}\right)^{-1}.
\ee
The optical depth 
$\tau=\tau_{\rm es}+\tau_{\rm abs}$,
with $\tau_{\rm es}=(1/2) k_{\rm es}\Sigma$ and
$\tau_{\rm abs}=Q_{\rm brem}/ (8\sigma T^4)$.
The equation of state is $p=p_{\rm gas}+p_{\rm rad}$, with
$p_{\rm gas}=\frac{\Re}{\mu}\rho T$, and
$p_{\rm rad}=\frac{Q_{\rm rad}}{4c}\left(\tau+\frac{2}{\sqrt{3}}\right)$
(Abramowicz et al. 1996). The bremsstrahlung radiation
$Q_{\rm brem}=2H(q_{ei}+q_{ee})$. The formula for $q_{ei}$ and $q_{ee}$
are from Narayan \& Yi (1995). Another useful equation is
the hydrostatic balance equation,
$H=c_s/\Omega_{\rm K}$, with $c_s=
(p/\rho)^{1/2}$ is the local sound speed.
The Paczy\'nski \& Wiita (1980) potential is used.

We solve  eqs. (2)-(7) to obtain the thermal equilibrium curve of LHAFs.
One noticeable parameter is $\xi$ in eq. (6). From \S1, we know
that for an LHAF, $Q_{\rm adv}<0$, 
therefore we should set $\xi < 0$ to recover this solution. 
We can recover all other accretion solutions by simply setting $\xi=1$, 
as in A95. The crudeness of the value of $\xi$
doesn't affect our qualitative results although it does prevent us from
obtaining the exact quantitative results such as the ranges of accretion rates
to which various accretion disk models correspond.

Figure 1 shows the thermal equilibrium curves of different solutions.
Comparing with the figures in A95 or C95, we see that the three 
``U'' shaped thick lines are new and they denote 
LHAFs with different values of $\xi$.
These lines extend from optically thin regime to optically thick regime,
bridging the SLE and the SSD solutions. 
It is not clear what physical implications the ``bridging'' has, or it might 
just be a mathematical trick. But we note in this context that 
in the slim disk solutions, they did find solutions with 
the advection term being negative in some cases (Abramowicz et al. 1988;
Szuszkiewicz, private communication). 
Some segment of the ``U'' curves are superimposed on the SLE
and SSD lines. We find that advection in these ``superimposed'' segment is
equal to zero, same with SLE and SSD.
Only the ``separate'' segment denotes the genuine LHAFs, since 
for this segment the energy advection is negative and 
plays a significant role in the energy balance
of ions. We see that above the
critical accretion rate of ADAF, $\dot{M}_1 (\sim 0.1\dot{M}_{\rm Edd})$,
both ADAF and SLE disappear and LHAF is the only available {\em hot} solution.
This is consistent with our analysis in \S 1.

For comparison, we also solve the {\em global} solutions of 
the standard accretion equations describing a 
one-temperature accretion flow with the same parameters as in Fig. 1
(see Yuan 1999 for the equations
except that the radiative term is now replaced by eq. (7) of the
present paper). The results are denoted by the filled circles (for ADAFs)
and triangles (for LHAFs). We can see that our local 
algebraic analysis is qualitatively consistent with the global results, i.e.,
when the accretion rate is lower than a certain value, the solutions are ADAFs;
while above this value, the solutions are LHAFs. We also calculated the 
value of $\xi$ for a specific example of $\dot{M}=10^{-1.6}\dot{M}_{\rm Edd}$
and the outer boundary conditions of $R_{\rm out}=10^3R_s,
T_{\rm out}=2 \times 10^9 $K, and $v/c_s=0.4$. We obtained $\xi \approx -0.3$
at $R=5R_s$. 

We want to emphasize that our ``U'' shaped line is completely different 
from the line with the similar shape in C95, which was obtained
by setting a very large viscous
parameter $\alpha > \alpha_{cr}$ with $ \alpha_{cr}\sim 0.2$ or
larger. In fact, a later more accurate treatment of microphysics and the
inner boundary condition by Bj\"ornsson et al. (1996)
indicated that $\alpha_{cr} > 1$, therefore, the ``U''
shaped branch in C95 is unphysical.

\section{Thermal stability}

From the density profile of the global solution in Y01, 
we know that LHAFs are viscously stable. An important problem
then is to analyze its thermal stability. Before doing that, we note
that an LHAF is dynamically very similar to a cooling flow in galactic clusters.
Many authors have studied the thermal stability of cooling flow
and concluded that it is 
unstable against local perturbations if thermal conduction is neglected
(e.g., Fabian \& Nulsen 1977; 
Mathews \& Bregman 1978; Nulsen 1986). Thermal conduction can strongly
stabilize the cooling flow (Zakamska \& Narayan 2003; Kim \& Narayan 2003).

Now let's investigate the thermal stability of LHAFs
against local perturbations. Such perturbations can be, e.g., 
that the local gas density increases wherever the initial 
ambient magnetic field is less than average. We follow the standard 
analysis approach presented by Kato, Abramowicz \& Chen (1996).
Taking the surface density as an example, we denote perturbations 
$\Sigma_1$ superimposed over the unperturbed quantities
$\Sigma$ as $\sigma \equiv \Sigma_1/\Sigma \propto 
{\rm exp}(n\Omega t-ikr)$ where $n\Omega$ is the growth 
rate of perturbation and $k$ is the perturbation wave number. 
Substituting the perturbed quantities to the time-dependent 
accretion equations, we can obtain the perturbed equations and
the dispersion relation. The arguments for ADAFs presented
in Kato et al. (1996)
hold for LHAFs as well although the advection is negative
and radiative cooling is important here.
We assume that thermal conductivity is suppressed by the
presence of the tangled magnetic field in the accretion flow (but see 
Narayan \& Medvedev 2001). We expect the flow will be thermally stable if thermal
conduction is strong. The dispersion relation is,
\be
-3n\Omega \sigma=G \sigma,
\ee
and the condition for instability is
\be
G \equiv 
-\frac{Q_{\rm vis}}{W}\left(\frac{\partial {\rm ln} \eta}
{\partial {\rm ln}\Sigma}\right)_{W}
-\frac{Q_{\rm rad}}{W}\left(\frac{\partial{\rm ln}Q_{\rm rad}}{\partial
{\rm ln}\Sigma}\right)_{W} < 0.
\ee
Here $W \equiv 2Hp$ and $\eta \equiv \nu \Sigma$. Since $\eta \propto 
\Sigma c_s^2/\Omega \propto W/\Omega$, so 
$(\partial {\rm ln} \eta/\partial {\rm ln}\Sigma)_W = 0$. 
For a one-temperature
LHAF, $Q_{\rm rad} \propto H \rho^2 T^{1/2} \propto \Sigma^2 T^{1/2}/H
\propto \Sigma^2$, so 
\be
(\partial{\rm ln}Q_{\rm rad}/\partial
{\rm ln}\Sigma)_W > 0. 
\ee
Therefore one-temperature LHAFs are thermally unstable 
against local perturbations.

The case of two-temperature LHAF is a little bit more complicated. 
Treating the energy
equation in Kato et al. (1996) as the energy equation of ions, the
instability condition is exactly the same as eq. (9) except that we should
replace $Q_{\rm rad}$ with $Q_{ie}$, the Coulomb collision cooling rate. 
Since $Q_{ie} \propto \rho^2 H T_e^{-3/2}
T_i$, to determine the sign of $G$, we must turn to the electron
energy equation to obtain the relation between $T_e$ and $T_i$.
Writing the energy equation 
of electrons as $Q_{\rm adv,e}=Q_{ie}-Q_{\rm rad,e}$, 
with $Q_{\rm adv,e}$
and $Q_{\rm rad,e}$ are the energy advection 
and radiative cooling of electrons respectively,
the global solution of Y01 indicates that
for LHAF a good approximation to the electrons
energy equation is $Q_{ie} \approx Q_{\rm adv,e}$ in the 
region outside of $\sim 100R_s$  and 
$Q_{ie} \approx Q_{\rm rad,e}$ inside $100R_s$ (see Figure 5 in Y01). 
Taking $Q_{ie} \approx Q_{\rm adv,e}$ we can get $T_e \propto 
W^{1/5}\Sigma^{1/5}$. Hence $Q_{ie} \propto
\Sigma^{6/5}W^{1/5}$ and
\be
G=-\frac{Q_{ie}}{W}\left(\frac{\partial{\rm ln}Q_{ie}}{\partial
{\rm ln}\Sigma}\right)_{W} ={6\over 5} > 0.
\ee
So a two-temperature
LHAF is thermally unstable at least at the region outside of $\sim 100R_s$. 
Inside $\sim 100R_s$, the radiative cooling
is in general dominated by the thermal 
Comptonization. The seed photons can be the synchrotron
photons, or other soft photons from cold matter 
such as a SSD underlying or outside of the 
LHAF, or cold clumps suspending in the hot gas. The 
origin of the cold clumps in LHAFs can be due to  the thermal instability
of LHAFs (see below), or originally existed in the
accretion material. For luminous sources the soft photons
from the cold matter is likely to dominate over the synchrotron photons 
as the seed photons of Comptonization (e.g., Zdziarski et al. 1998).
Since $Q_{\rm rad,e} =(4kT_e/m_ec^2)\rho H 
\kappa_{es}U_r c$, with $U_r$ denotes
the flux of seed photon, using $Q_{ie}\approx Q_{\rm rad,e}$,
we find that if $U_r \propto T_i^2$ or stepper, the two-temperature
LHAFs will be thermally stable, otherwise it is unstable. 
Unfortunately, it is not easy to determine whether this condition
can be satisfied or not, because of the complicated energy interaction 
between the cold and hot phases in accretion flow (e.g., Ferland \& Rees 1988).

Now let's discuss the consequence of the 
thermal instability against local perturbations.
An important factor is how fast  
the perturbations grow compared to other related timescales.
It is believed that as the result of instability cold dense clumps will form
in the hot flow if the growth of perturbations is 
fast enough. This mechanism has been used
to explain the condensations of galaxies from the intergalactic
medium, formation of solar prominence, and condensations
in planetary nebulae (e.g., Field 1965). 
In the case of accretion flows, we should compare the 
accretion timescale $t_{\rm acc}$ with the growth timescale 
of perturbations $t_{\rm grow}$.
For LHAFs, from eq. (8), the timescale of the growth of 
local perturbations is
\be
t_{\rm grow}=(n\Omega)^{-1}=3 G^{-1}=3 \left[\frac{Q_{\rm rad}}{W}
\left(\frac{\partial{\rm ln}Q_{\rm rad}}{\partial
{\rm ln}\Sigma}\right)_{W}\right]^{-1}=\frac{3 W}{2Q_{\rm rad}}
\ee
This equals the thermal timescale of accretion flows, $t_{\rm th}$, 
\be
t_{\rm grow}=\frac{3(\gamma-1)}{2}t_{\rm th}=t_{\rm th},
\ee
for adiabatic index $\gamma=5/3$. 

We compare the two timescales, $t_{\rm acc}$ and 
$t_{\rm grow}$, by numerical calculations. This requires us to solve
the global solution of two-temperature LHAFs.
This was done in Y01. The results indicate that
depending on the value of $\dot{M}$, there are two types of LHAF.
When $\dot{M}_1 \la \dot{M} \la \dot{M}_2$,
with $\dot{M}_2$ can be as high as $5\dot{M}_1$,
the accretion gas in LHAF is hot throughout the disk;
when $\dot{M} \ga \dot{M}_2$, up to $\dot{M}_{\rm Edd}$,
within a certain radius the density of the accretion gas is so high that even
the sum of compression work and viscous dissipation can not balance the strong
Coulomb cooling. As a result, the hot accretion flow will collapse onto
the equatorial plane and form an optically thick cold annulus.
While the quantitative results such as the value of $\dot{M}_2$ depend on 
parameters such as $\alpha$, the qualitative results should not.

Correspondingly, there are two kinds of results when we
compare $t_{\rm acc}$ and $t_{\rm grow}$, as shown by the two lines in Figure 2.
The corresponding two models are taken from Fig. 4 of Y01.
The solid line corresponds to $\dot{M}=0.1\dot{M}_{\rm Edd} < \dot{M}_2$, 
so the accretion flow is hot throughout the disk; 
the dashed line to $\dot{M}=0.3\dot{M}_{\rm Edd} 
> \dot{M}_2$, so the hot accretion flow collapses at a radius and 
form a cold annulus. We see that when $\dot{M}< \dot{M}_2$,
$t_{\rm grow}/t_{\rm acc} > 1$, so the perturbations have
no time to grow before they are swallowed by the black hole,
therefore, the thermal instability will have no effect on the dynamics of LHAFs.
When $\dot{M}> \dot{M}_2$, $t_{\rm grow}/t_{\rm acc} < 1$ at the transition radius.
As a result, some cold dense 
clumps will form there if LHAFs are thermally unstable at that region. 
We speculate that once the condensation of the hot phase makes
$t_{\rm grow}\approx t_{\rm acc}$, the condensation will stop.
This constraints will be useful when 
quantitatively calculating
the physical states of clumps such as their filling factor, density
and temperature.

In our calculations of $t_{\rm acc}$ and $t_{\rm grow}$,
we considered only the ``standard'' radiative processes, namely
synchrotron, bremsstrahlung, and their Comptonization.
If we include the additional Compton cooling due to the soft photons from
the cold phase, $t_{\rm grow}$ will decrease (
$\dot{M}_1$ and $\dot{M}_2$ will also decrease). We expect  
that we will still have $t_{\rm grow} > t_{\rm acc}$ when $\dot{M}<\dot{M}_2$.

Note that LHAF will not collapse into an optically thick
cold disk as the result of {\em local} perturbation since the perturbation
wavelength is smaller than the disk scale-height.
The thermal stability of LHAFs against long wavelength perturbations is hard to
determine, and it is also not clear how fast the growth of 
perturbations is compared to the accretion timescale if the flow is unstable.
On the other hand, if the filling factor of the cold clumps
is large enough, it is possible that these clumps
may assemble due to frequent collisions and form a disk-like
large scale structure.

Many authors have proposed such cold/hot two-phase accretion
flow model for AGNs (e.g., Guilbert \& Rees 1988;
Ferland \& Rees 1988; 
Kuncic, Celleti \& Rees 1997). But almost all these work 
focus on the thermal state
of cold clumps and how the clumps re-radiate the energy they absorb.
Krolik (1998) also suggested a two-phase accretion flow model.
Different from our suggested physical mechanism of clumps formation,
he suggested that such a two-phase feature is the result
of the instability in the radiation-pressure-dominated innermost
region of a SSD (see e.g., Gammie 1998).


\begin{acknowledgements}
I thank Shoji Kato, Andrei Beloborodov, and Ramesh Narayan
for valuable discussions, and the anonymous referee for helpful
comments. This work was supported in part by NASA grant NAG5-10780 and
NSF grant AST 0307433.
\end{acknowledgements}

{} 

\clearpage

\begin{figure}
\plotone{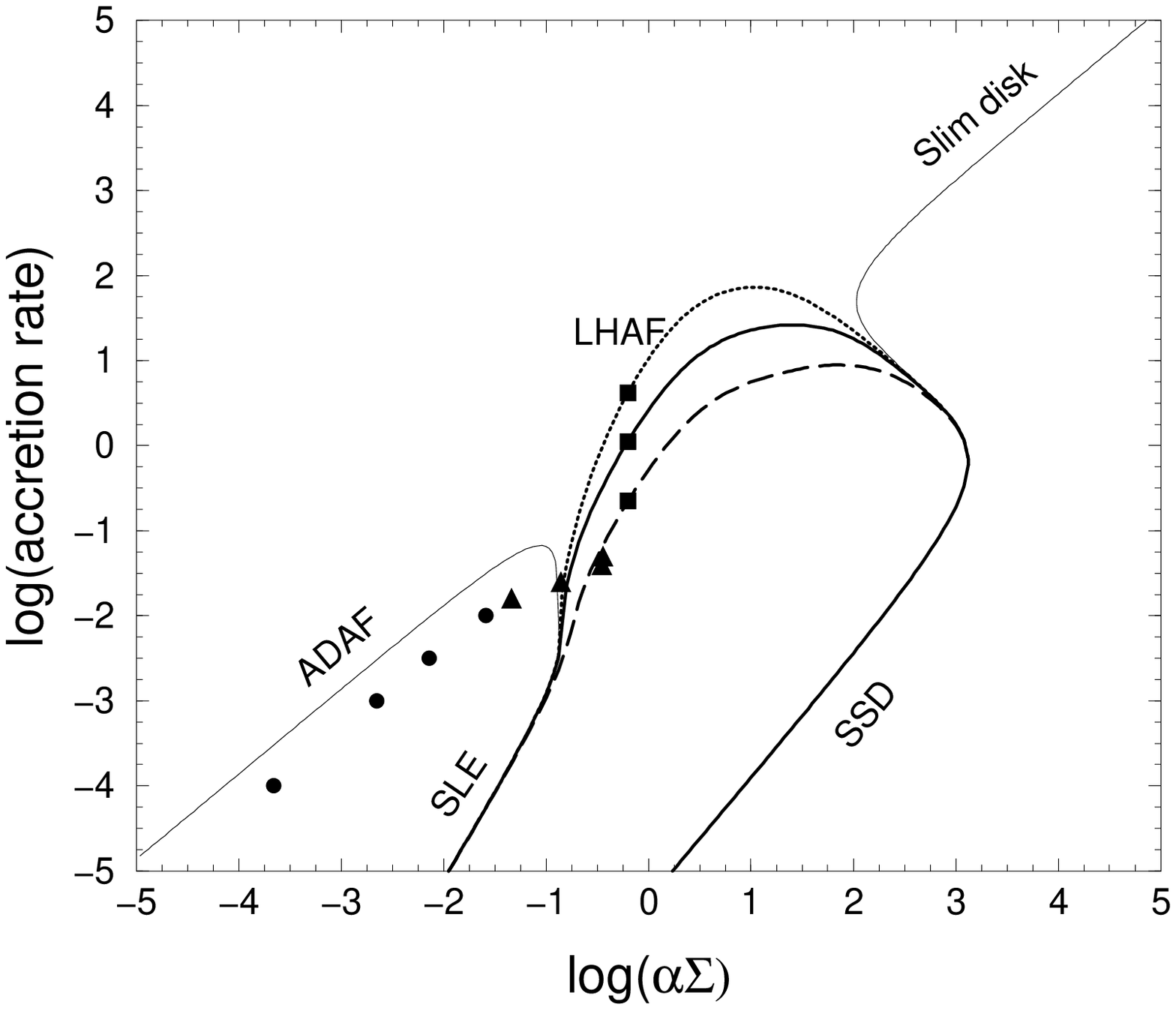}
\caption{The thermal equilibrium curve of various accretion solutions.
The accretion rate is in units of $\dot{M}_{\rm Edd}\equiv 10L_{\rm Edd}/c^2$
and the units of $\Sigma$ is g~cm$^{-2}$.
The parameters are $M/\msun=10,
\alpha=0.1$, and $R=5R_s$. The thin solid lines are for $\xi=1$,
representing ADAF, SLE, SSD, and
slim disk. The thick solid, dotted and dashed lines
are for $\xi=-1, -0.1$ 
and $-10$, respectively, representing LHAFs. 
The squares are the location of unity scattering optical depth.
The filled circles and triangles are the results obtained from the global
solutions of ADAFs and LHAFs, respectively.}
\end{figure} 

\begin{figure}
\plotone{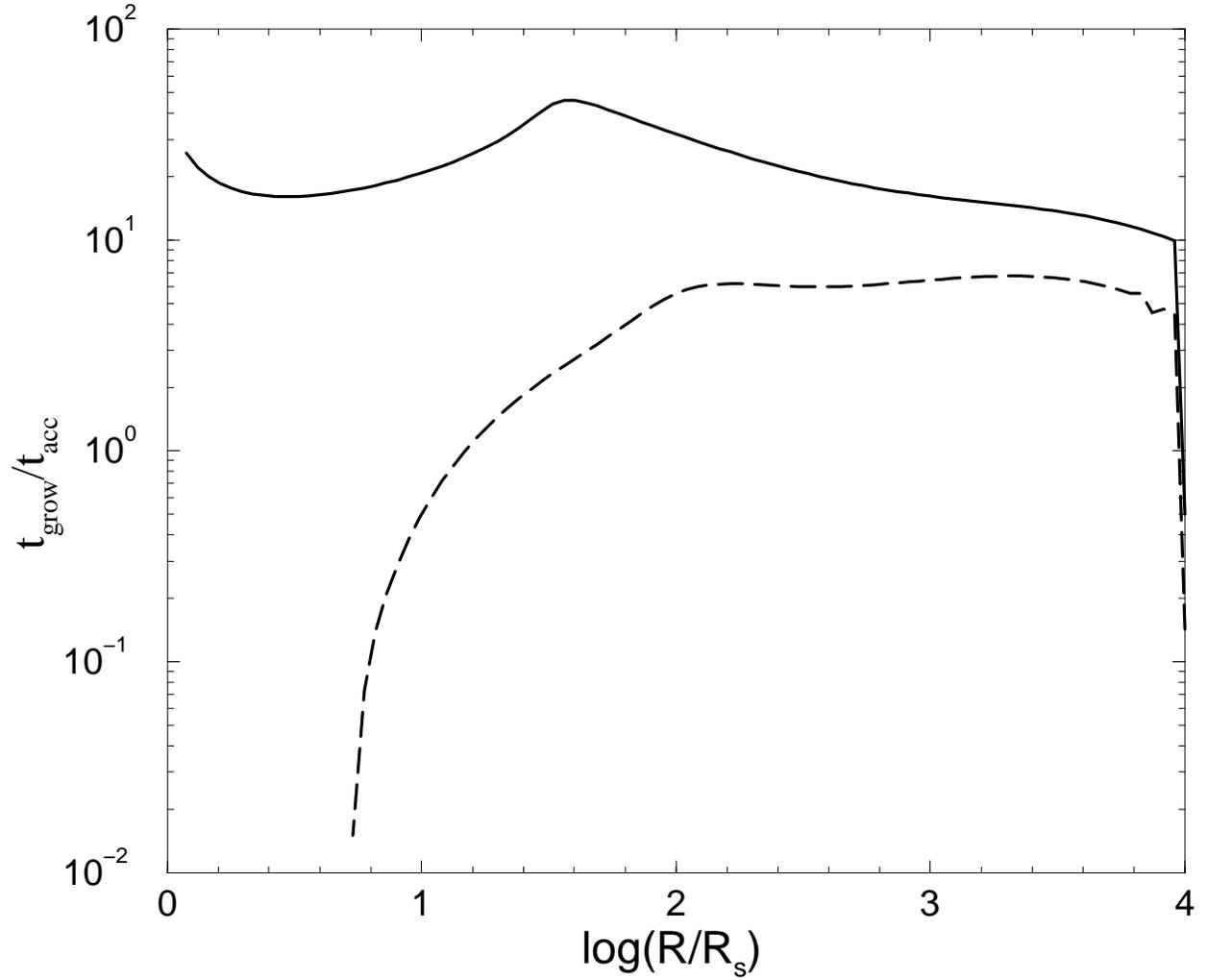}
\caption{The ratio of the accretion timescale 
to the timescale of growth of perturbations as a function of radii for
two LHAF solutions presented in Fig. 4 of Y01. The solid (dashed) line
is for $\dot{M}=0.1 (0.3)\dot{M}_{\rm Edd}$. 
Other parameters are $\alpha=0.1, \beta=0.5$.}
\end{figure}


\end{document}